\documentclass{nature}

\usepackage{aas_macros}
\usepackage{amssymb}
\usepackage{graphicx}
\usepackage{color}
\usepackage[normalem]{ulem} 
\usepackage{soul}
\usepackage{url}
\usepackage[colorlinks]{hyperref}

\usepackage{subfig}
\usepackage[font=footnotesize]{caption}

\usepackage{floatrow}
\DeclareFloatFont{tiny}{\scriptsize}
\title{Confinement of relativistic particles in the vicinity of  accelerators: a key for understanding the  anomalies in  secondary cosmic rays}

\begin{document}

\maketitle

\author{
Ruizhi Yang$^{1,2,3,4}$,
Felix~Aharonian$^{4,5,6,7}$,
}

\begin{affiliations}
\small
\item CAS Key Laboratory for Research in Galaxies and Cosmology, Department of Astronomy, School of Physical Sciences,University of Science and Technology of China, Hefei, Anhui 230026, China
\item Deep Space Exploration Laboratory, Hefei 230026, China
\item School of Astronomy and Space Science, University of Science and Technology of China, Hefei, Anhui 230026, China 
\item Tianfu Cosmic Ray Research Center, 610000 Chengdu, Sichuan, China
\item Dublin Institute for Advanced Studies, 31 Fitzwilliam Place, Dublin 2, Ireland 
\item Max-Planck-Institut f\"ur Kernphysik, P.O. Box 103980, D 69029 Heidelberg, Germany 
\item Gran Sasso Science Institute, 7 viale Francesco Crispi, 67100 L'Aquila,  Italy

\end{affiliations}

\hfill

\begin{abstract}
Recent cosmic ray (CR) measurements have revealed unexpected anomalies in secondary CRs, namely deviations from the predictions of the so-called standard Galactic CR paradigm regarding the composition and energy spectra of the products of interactions of primary (accelerated) CRs with interstellar gas: (i) antiparticles (positrons and antiprotons), (ii) light elements of the (Li, Be, B) group, and (iii) diffuse gamma rays. We argue that the new measurements can still be explained within the standard CR paradigm but with an additional assumption that CRs spend a significant part of their lifetime near their formation sites. The latter can be realized if CRs propagate more slowly in these localized regions than in the interstellar medium (ISM). Postulating that CRs accumulate on average energy-independent "grammage" of $0.7 \ \rm g/cm^2$ near the major contributors to galactic CRs,  one can explain self-consistently the new measurements of the B/C ratio by DAMPE and the diffuse ultra-high-energy gamma-rays by LHAASO,  involving a minimal number of model parameters: the energy-dependent "grammage" in the interstellar medium 
$\rm \lambda \approx 8 (E/10 \ GeV)^{-0.55}~\rm g/cm^{2}$  and the average CR
acceleration (sourcee) spectrum \, $\rm Q(E)  \propto E^{-2.3}$. 
\end{abstract}
\newpage

Relativistic particles accelerated in Galactic Cosmic Ray Factories and propagating through interstellar magnetic fields represent a fundamental component of the Milky Way. According to the current Galactic CR paradigm, protons and nuclei are accelerated, up to and beyond the so-called {\it knee} around a few PeV, in galactic sources, and confined in the halo surrounding the Milky Way (MW). During the propagation through the interstellar magnetic fields, they interact with the ambient gas and produce the so-called secondary CR component.  The latter contains unique information about the history and dynamics of the production and confinement of CRs in the Galaxy. 
The secondary CRs, first of all, positrons, antiprotons, and the nuclei of the group of (Li, Be, B), which are almost missing in the ambient matter, give us information about the energy-dependent "grammage" or the confinement time.  On the other hand, the diffuse (secondary gamma-rays tell us about the density and spatial distribution of CRs throughout the galactic plane. 

If  CRs are injected with a rate $Q(E)$,  the steady-state distribution of particles established in ISM is determined by the CR confinement time  $\tau(E)$ in the Galaxy, $N(E) \propto Q(E)\tau(E)$. Here, it is assumed that the energy losses due to interaction with the ambient medium are negligible. This is the case for CR protons and nuclei but not for the high-energy electrons and positrons, whose energy losses through the synchrotron and inverse Compton cooling in the TeV band dominate over the escape. Generally, the confinement time decreases with energy as $\tau(E) \propto  E^{-\delta}$.  Thus, for a power-law injection spectrum, $Q(E) \propto E^{-\gamma_{}}$,  we have $N(E) \propto E^{-(\gamma_{}+\delta)}$.  Ignoring the slight energy dependence of the inelastic cross-sections above 10~GeV, the production spectra of secondaries mimic the steady-state spectrum of primaries in the ISM. Due to the energy-dependent propagation, which obviously should be described by the same diffusion coefficient for both primary and secondary CRs, the steady-state spectrum of secondaries can be written as  $N_{\rm sec}(E) \propto Q(E) \tau^{2}(E) \propto  E^{-(\gamma_{}+2\delta)}$. Consequently, the  secondary-to-primary ratio should decrease monotonically with energy,  $R(E)=N_{\rm sec}/N \propto E^{-\delta}$. This simple scenario at low energies agrees well with the measured content of the secondary light nuclei \cite{ams02bc}. 


\section{Results}
However, the recent progress in extending the direct CR measurements to higher energies revealed several `anomalies' in the spectrum of secondary CRs, such as the surprisingly similar spectral shapes of positrons and antiprotons as that of the primary CR protons\cite{ams_antiproton,ams02posi}, thus challenging the standard CR paradigm. In our previous work \cite{yang19}, we argued that accumulating a non-negligible 'grammage' in or near the particle accelerators can potentially address this unexpected result. We, in particular, predicted an increase in the Boron to Carbon (B/C) ratio at higher energies. Recently, DAMPE  collaboration has extended the B/c measurements up to the kinetic energy of $\sim 4~\rm TeV$ per nucleon\cite{dampe_bc}, revealing significant hardening of the B/C ratio, in contrast to the prediction of the standard CR paradigm:  $\rm R(E)=B/C \propto E^{-\delta}$.  Instead, these data agree well with the predictions of Ref. \cite{yang19}. 

Based on the new B/C measurements, we modified the parameters used in ref.\cite{yang19}. By fixing the average grammage near the CR sources to about $0.7 ~\rm g/cm^2$, and assuming its energy independence, at least up to $10~\rm TeV$, one can satisfactory fit the  DAMPE results (see Fig.1). The increase of the B/C ratio is explained by the contribution of the secondary component produced near the accelerators, which, above a few hundred GeV, begins to dominate over the component made in ISM. 


\begin{figure}
\centering
\includegraphics[scale=0.6,angle=-90]{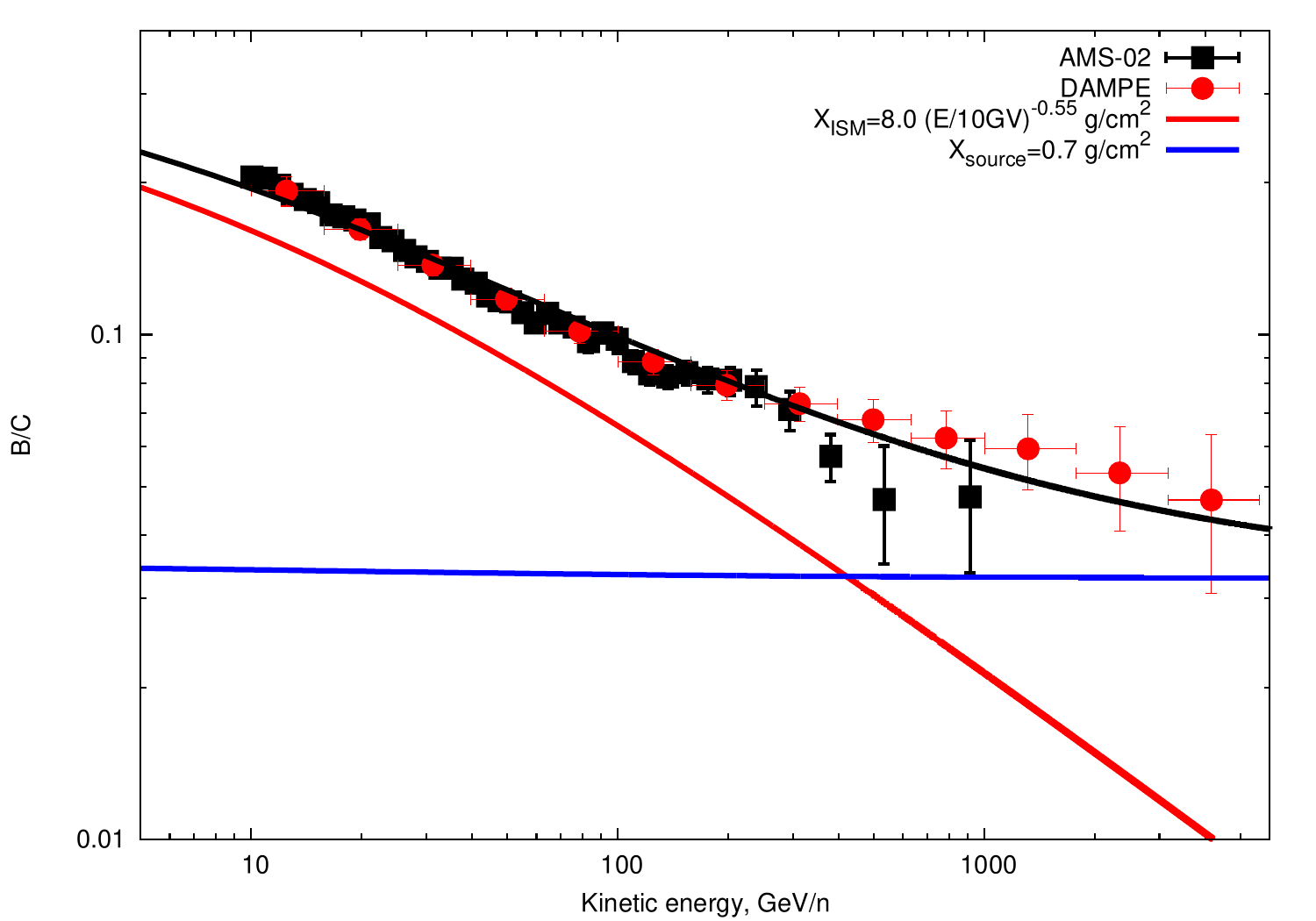}
\caption {
The AMS and DAMPE measurements of the boron-to-carbon ratio.  
The theoretical curve is calculated within the modified CR paradigm, assuming the secondary boron nuclei are comparably produced near the CR accelerators and in the interstellar medium. The parameters are described in the text.
}
\label{fig:bc}
\end{figure}

 \begin{figure}
\centering
\includegraphics[scale=1.2]{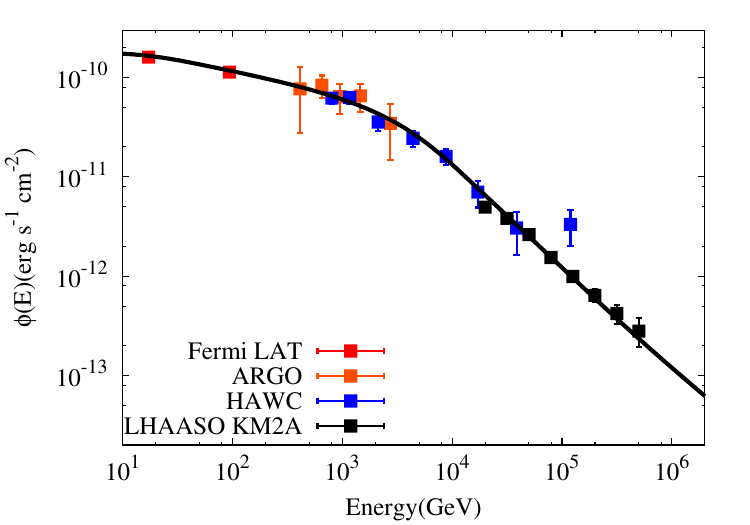}
\caption {
The broadband gamma-ray spectrum of Cygnus region. The curve presents the gamma-ray spectrum produced by parent CR protons with a broken power law, the index is 2.3 and 3.0
}
\label{fig:spe}
\end{figure}


If in proximity to their production sites CRs propagate slower than in ISM, they can accumulate non-negligible grammage and thus proficiently produce secondary particles \cite{cowsik15,bresci19,recchia22,sun23,nie24}. This cannot be excluded since, in these regions, the environment can be highly turbulent with an enhanced magnetic field. Moreover, an active CR accelerator may initiate in the surrounding regions favorable conditions for a slow, self-regulated collective CR escape  \cite{malkov13}. 

The effective confinement around the  CR accelerators results in enhanced CR density, which can significantly exceed, depending on the CR injection power and the diffusion coefficient, the density of the locally measured CRs (the 'cosmic ray sea') up to distances of 100~pc or more.   This would substantially increase the $\gamma$-ray production emissivity, forming extended diffuse gamma-ray structures surrounding the accelerators. 

A large fraction of extended galactic $\gamma$-ray sources reported by LHAASO  with spectra extending to 100 TeV and beyond\cite{lhaaso_catalog} can result from this scenario.  
In this category of sources ("PeVatron Halos"), a particular interest is the so-called Cygnus  Bubble, a diffuse UHE source detected in the direction of the star-forming region Cygnus X \cite{lhaaso_cygnus}. The extension of the energy spectrum out of 1~PeV  \cite{lhaaso_cygnus}  implies the presence of a hadronic Super-PeVatron(s) somewhere with an angular area exceeding $\rm 100 \ deg^2$. Moreover, based on the gamma-ray morphology and the gas distribution, the LHAASO collaboration has derived the radial distribution of CR protons inside approximately 100~pc radius. It appeared close to $1/r$, implying that the CR accelerator(s) is presumably
located in the Stellar Cluster Cyg OB2 \cite{lhaaso_cygnus}. 

The overall gamma-ray spectrum of Cygnus Bubble is measured with good accuracy from 1 Gev to 1PeV. It can be described by a broken-power law with photon indices $\alpha=2.2$ and 3.0 below and above $\rm E_{\rm br} \sim 2~\rm TeV$, respectively. This radiation spectrum can be explained by $\pi^0$-decay gamma-rays from a parent proton population with a broken power-law spectrum: $\gamma = 2.3$ and 3.1  below and above $20~\rm TeV$, respectively (see Fig. \ref{fig:spe}). Such proton spectrum can be naturally explained by a power law proton injection spectrum with $\gamma=2.3$, together with an energy-independent diffusion coefficient below $\sim 20 ~\rm TeV$,  
but $D(E) \propto  E^{0.8}$ above,  
close to the phenomenological model parameters applied in ref.\cite{lhaaso_cygnus}.

For  the CR radial distribution $\rm w(r) \propto  r^{-1}$, the total proton energy within the volume of radius  $R$  is  \cite{aharonian19}
\begin{equation} 
W_{\rm  p}= 4 \pi \int_0^{R_{\rm 0}}  w(r) r^2  \,\mathrm{d}r  \approx \\
 2.7 \times 10^{47} \frac{w_0}{1 \ \rm eV/cm^3} (\frac{R}{10 \ \rm pc})^2 \ \rm erg \ ,
 \label{Wcr}
 \end{equation} 
 While the energy density of CRs $w_0$  at 10~pc  can be found from the gamma-ray emissivity measured directly from observations,  the source's radius may contain considerable uncertainty.  Because of the limited instrument sensitivity to low brightness, the region filled by CRs from the central source can be much larger than the size of the revealed diffuse gamma-ray source. 
The integration radius $R_0$ 
can be estimated theoretically, assuming that it represents the diffusion radius:    $R_D = 2 \sqrt{T D(E)}  \approx   3.6  \times 10^3  (D_{30}  T_{6})^{1/2} \  \rm pc$, where  $ D_{30}$  is the diffusion coefficient of protons in the units of  $10^{30} \rm cm^2/s$, and $T_{6}$ is the age of the CR accelerator normalized to $10^6$ years. Assuming that the CR accelerator is linked, in one way or another, to Cyg OB2, with age estimated between 1 and 7 times $10^{6} \rm years$ \cite{wright15}, and using for CR energy density above 10~TeV  the estimate  $w_0 \sim 0.5~\rm eV $ as derived in ref. \cite{lhaaso_cygnus}, we have

\begin{equation} 
W_{\rm  p}(E>10 \, \rm PeV) \approx 1.8 \times 10^{52} \, D_{30} \, T_{6} \ \rm erg \ ,
 \label{Wcr1}
 \end{equation} 
 
 On the other hand, the total energy in CRs injected over the lifetime of the source, $T_0$, can be expressed through the available energy contained in the stellar winds in Cygnus~OB2:
\begin{equation}
W_{\rm  tot}=f L_0 T_0\approx3 \times 10^{52}f  L_{39} T_6 \ \rm erg, 
 \label{W_CR}
 \end{equation}
where $L_{39}=10^{39} L_0$ is the total mechanical power of the stellar winds in units of $10^{39} \ \rm erg/s$, and $f$ is the efficiency of conversion of the wind kinetic energy to CR protons with energy larger than 10 TeV.  Equating Eq.(\ref{Wcr1}) and  Eq.(\ref{W_CR}),  we find 
\begin{equation}
 D_{30}\approx1.7 f L_{39}  .
 \label{f_D} 
 \end{equation}
Since the acceleration efficiency of protons with energy $>10~\rm TeV$ can hardly exceed $1\%$, Eq.\ref{f_D} gives an upper limit on the diffusion coefficient $D \simeq 10^{28}~\rm cm^2/s$ which is smaller, by two orders of magnitude, than the relevant value in the Galactic Disk, $D_{30} \sim 1$ \cite{strong07}. 


The slower diffusion of CRs near their accelerators has been theoretically predicted and motivated by the enhancement of the magnetic turbulence caused by CRs \cite{malkov13,dangelo18}. In particular, the so-called CR streaming instability can generate such high turbulence. In this case,   the diffusion coefficient is expected to be energy-independent up to a specific energy $E^*$, after which it increases with energy\cite{krumholz20}. The latter should lead to the steepening of the CR spectrum established inside the bubble, $f_{\rm CR} \propto Q(E)/D(E)$, above $E^*$. Correspondingly, one should expect a break in the spectrum of secondary gamma rays at $\sim 1/10 E^*$.

Fig. 2 shows the energy spectrum of Cygnus Bubble above 10~GeV. 
%
In this energy range, the spectrum of $\pi^0$-decay gamma rays is slightly harder than the spectrum
of parent protons and is shifted toward lower energies by a factor of 10. Despite the break, the 
gamma-ray spectrum can be explained by a pure power-law injection spectrum of protons, $Q(E) \propto E^{-2.3}$,  but combined with energy-independent diffusion coefficient up to proton energy $\approx 10$~TeV, and $D(E) \propto E^{-0.7}$ above it. 

Similar extended UHE gamma-ray structures have also been detected around other stellar clusters  \cite{hess_w1, yang18, yang20w43}. 
Assuming, for simplicity, that Cygnus Bubble is an average representative of this source population,  the 
grammage accumulated by CRs inside the Bubble can be estimated as $\chi \sim n m_p c T_{\rm c}$  where $n$ is the number density of the ambient gas, and $T$ is the confinement time, and $c$ is the speed of light. If CRs escape the bubble diffusively, their confinement time is estimated as $T_{\rm c} \simeq R_{\rm c}^2/4D$, where $R_{\rm c}$ is the region where CRs are confined. We can take the {\it resolved} linear size of the gamma-ray emitter as the lower limit for the latter. The source could be more extended, i.e., the confinement radius could be larger. Indeed, the outskirts of the source could not be detected because of the reduced brightness below the detector's sensitivity.  
Taking the size of the gamma-ray emission region $R_{\rm c}$ of about  100~pc, the gas density $n\sim 1 ~\rm cm^{-3}$ and the diffusion coefficient below $\approx 10$~TeV of about $D \sim 10^{27}~\rm cm^2/s$,  the grammage accumulated in the bubble can be estimated as $\chi \sim 1  ~\rm g/cm^2$. 

The diffusion coefficient and the gas density cannot be identical in different sources; moreover, they can be distributed inhomogeneously inside the bubbles. Yet, the required combination of three parameters - the power-law index of the CR injection spectrum 
$\Gamma_0 \approx 2.3$, the energy-independent diffusion coefficient below $\sim 10$~TeV, and the corresponding grammage $\approx 1 \ \rm g/cm^2$ -   match well the set of the parameters used in Fig. 1 for the explanation of the B/C ratio of secondary CRs.  



In Fig.1, for simplicity,  we assume that all CRs accumulate a grammage of  $0.7 ~\rm g/cm^2$ near their accelerators before entering into ISM. Due to the negligible energy losses and the effective mixture of CRs in ISM, this estimate should be regarded as an average grammage near the "standard" CR source.    
Although we cannot fix the parameters responsible for the secondary nuclei and antiparticles (positrons and antiprotons) for {\it individual sources}, the observations of secondary gamma-rays produced in the same interactions allow us to study the scenario on a "source-by-source" basis by detecting giant diffuse gamma-ray halos surrounding the major CR accelerators in the Milky Way. 

\begin{figure}
\centering
\includegraphics[scale=0.6,angle=-90]{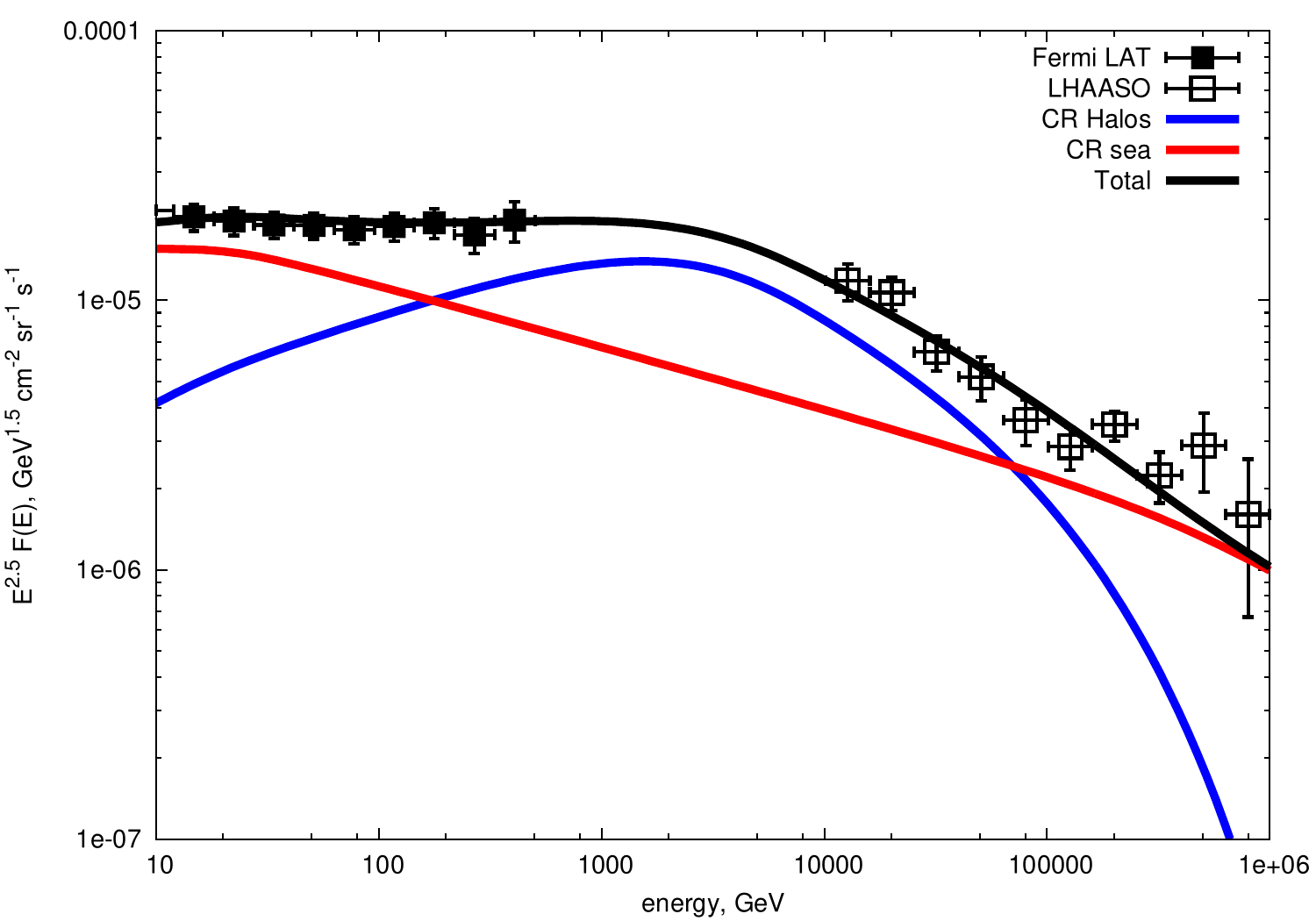}

\caption {
The diffuse emission measured by Fermi LAT and LHAASO KM2A in the inner Galaxy \cite{lhaaso_diffuse,zhang23}. The curves show the diffuse emission from the interaction of the 'CR sea' with the ISM (`BKG') assuming the p[position 
of the CR proton `knee' at  $2~\rm PeV$, as well as the gamma-ray emission from the slow diffusion regions near the sources (`CR bubbles'), assuming that the gamma-ray spectra in all 'CR bubbles' are the same as that in Cygnus Bubble \cite{lhaaso_cygnus} below 100 TeV calculated for the proton spectrum with an exponential cutoff 
at $1~\rm PeV$.  
 }
\label{fig:dif}
\end{figure}

An important implication of this scenario, with very slow diffusion in the regions near the CR accelerators, is the non-negligible, if not the dominant, contribution to the diffuse gamma-ray emission
\cite{yang19}. One should note that the LHAASO collaboration has detected a significant excess in the TeV-to-PeV gamma-ray diffuse emission in the Galactic plane\cite{lhaaso_diffuse}. The excess in the galactic plane at lower energy was also detected by Fermi-LAT \cite{fermi_diffuse, yang16, zhang23}.  
The extra contribution to diffuse gamma-ray emission by CR Bubbles provides a natural explanation for this excess. In Fig.\ref{fig:dif}, we plot the diffuse gamma-ray emission towards the inner Galaxy measured by LHAASO and FERMI LAT. The superposition of the component originating from CRs globally in ISM (labeled as "ISM"), together with the contribution from slow diffusion regions (labeled as "CR Bubbles"), give a satisfactory fit to the observed data. The "ISM" component is the truly diffuse gamma-ray emission produced in the Galactic Disk by the "Cosmic Ray "sea". This component has been studied extensively and can be modeled with the universal CR propagation models such as GALPROP \cite{galprop}  and DRAGON \cite{dragon}.

Remarkably, the contributions to the diffuse $\gamma$-ray emission from giant CR Bubbles and the "ISM" component can be derived directly from the measurements of the B/C ratio (Fig.\ref{fig:bc}) without any normalization. Indeed, the $\gamma$-ray flux $F_{\gamma} \sim n_{\rm CR}n_{\rm gas}$, where $n_{\rm CR}$ and $n_{\rm gas}$ is the CR and gas densities, respectively. The CR density, 
$n_{\rm CR} \sim Q_{\rm CR}\tau$, where $Q_{\rm CR}$ and $\tau$ are the CR injection rate and the confinement time, respectively. In Fig.\ref{fig:bc}, the grammage can be expressed as $\chi \sim n_{\rm gas} \tau $. 
Consequently, the relative fractions to the grammage contributed by these two components can determine the corresponding $\gamma$-ray fluxes. 
The only difference is that in spallation reactions, the kinetic energy per nucleon keeps constant, while in the $\gamma$-ray production, the average $\gamma$-ray energy scales as $1/10$ of the parent CR proton energy. 

As shown in Fig.\ref{fig:bc}, these two components have a comparable contribution at the rigidity of about $1~\rm TeV$.  
Thus, the $\gamma$-ray flux from the "bubble" component should begin to dominate at the energy of about $100~\rm GeV$. 
The calculations using the $\gamma$-ray production cross sections compiled in ref.\cite{kafexhiu14} are shown in Fig.\ref{fig:dif}. 
For the "ISM" component, we use a power-law CR spectrum with a power-law index 2.8. For the "bubble" component, we use a broken power law spectrum of parent CRs with an index of 2.3 below $15~\rm TeV$ and 3.0 above, with an exponential cutoff at $1~\rm PeV$. The relative normalization comes from the grammage of the two components as shown in Fig.\ref{fig:bc}. 
Such a setup can satisfactorily fit the observed diffuse $\gamma$-ray emission in the Fermi-LAT and LHAASO KM2A band. Because of the inhomogeneity of both the CR and gas distributions and the anticipated differences of the relevant physical parameters in different "CR bubbles," these calculations should be regarded as the first-order approximation. Yet, they already give a surprisingly accurate description of observed data. The only additional parameter introduced in calculations is the energy cutoff in the spectrum of injected CRs $E_{\rm 0} =1~\rm PeV$.  Although in Cygnus Bubble, the cutoff should be well above 1~PeV,  this is a reasonable assumption since the ultra-high energy signals so far are detected only from Cygnus Bubble, and we do not expect that all CR accelerators operate as Super PeVatrons. The fit to the highest energy data points also depends on the contribution of the "ISM" component in the `knee' region, whose spectrum presently is not well constrained.

Another prediction of the suggested interpretation of the diffuse galactic emission as a superposition of contributions of the "ISM" and "CR Bubbles" components concerns the change of the diffuse fluxes in different directions. 
At low energies,  below several hundred GeV, the gamma-ray diffuse emission mainly comes from the `ISM' component.  For a relatively homogeneous distribution of CRs in ISM, this component should correlate with the gas distribution.  
At intermediate energies between $1~\rm TeV$ and  $30~\rm TeV$,  the gamma-ray diffuse emission is dominated by the `CR bubbles' component. Thus, it will follow the distribution of CR sources more closely. Due to the intermittency and discreteness of CR sources, the gamma-ray diffuse emission could reveal significant variations in different directions, independent of the gas column densities.  Future LHAASO measurements, after substantial 
enhancement of the photon statistics, can potentially test this prediction.


\begin{figure*}
\centering
\includegraphics[scale=1.2,angle=0]{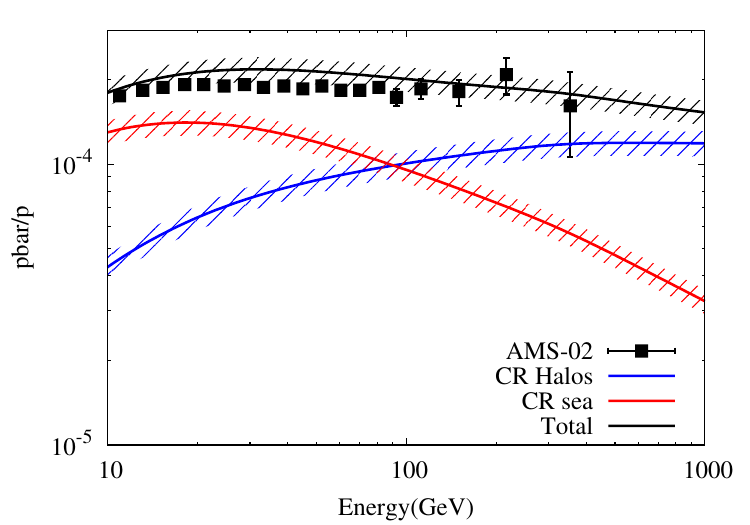}
\caption {
The AMS-02 measurement of the antiproton/proton ratio.  
The theoretical curve is calculated using the same grammage as in Fig.\ref{fig:bc}. The shaded area marks the uncertainties of the cross sections. 
}
\label{fig:ap}
\end{figure*}

\section{conclusion}

To conclude, the recent  CR measurements challenge the traditional paradigms regarding the acceleration and propagation of CRs in the Milky Way. The unexpected spectral hardening observed in the B/C ratio up to  $\approx 4$~TeV per nucleon suggests the presence of another component of CR secondaries. On the other hand, the detection of extended gamma-ray structures near potential CR accelerators indicates slower diffusion and effective CR confinement in these regions, accumulating large grammage by CRs. This results in an additional non-negligible component of secondary CRs. Furthermore, the excesses observed in diffuse gamma-ray emission in the Galactic plane may be attributed to the contributions from specific slow diffusion regions  - 
hadronic gamma-ray halos surrounding CR accelerators. 

In our model,  other secondary CR species, such as positrons and antiprotons, 
are unavoidably produced in the same structures. This may be responsible for the positron excess revealed by the direct measurement in the solar system \cite{ams02posi,pamelae}. However, high energy CR positrons suffer severe radiative cooling; therefore, the detected positron fluxes are presumably dominated by a single or a few nearby sources \cite{AAV2}. This makes CR positrons essentially different from the heavy secondaries (antiprotons and light nuclei). Thus, the above approach cannot be directly applied to CR positrons.  The situation is different for antiprotons; they can be treated as the secondary CR nuclei.  The antiproton/proton ratio, calculated 
assuming the same model parameters as for the B/C ratio and using the parametrization of cross sections from ref.\cite{winkler18}, is shown in Fig.\ref{fig:ap}.   One can see that within the uncertainties in relevant cross sections (about 10\%) \cite{korsmeier18}, the calculations are consistent with the AMS-02 antiproton measurements \cite{ams_antiproton}.

\begin{figure}
\centering
\includegraphics[scale=0.6,angle=-90]{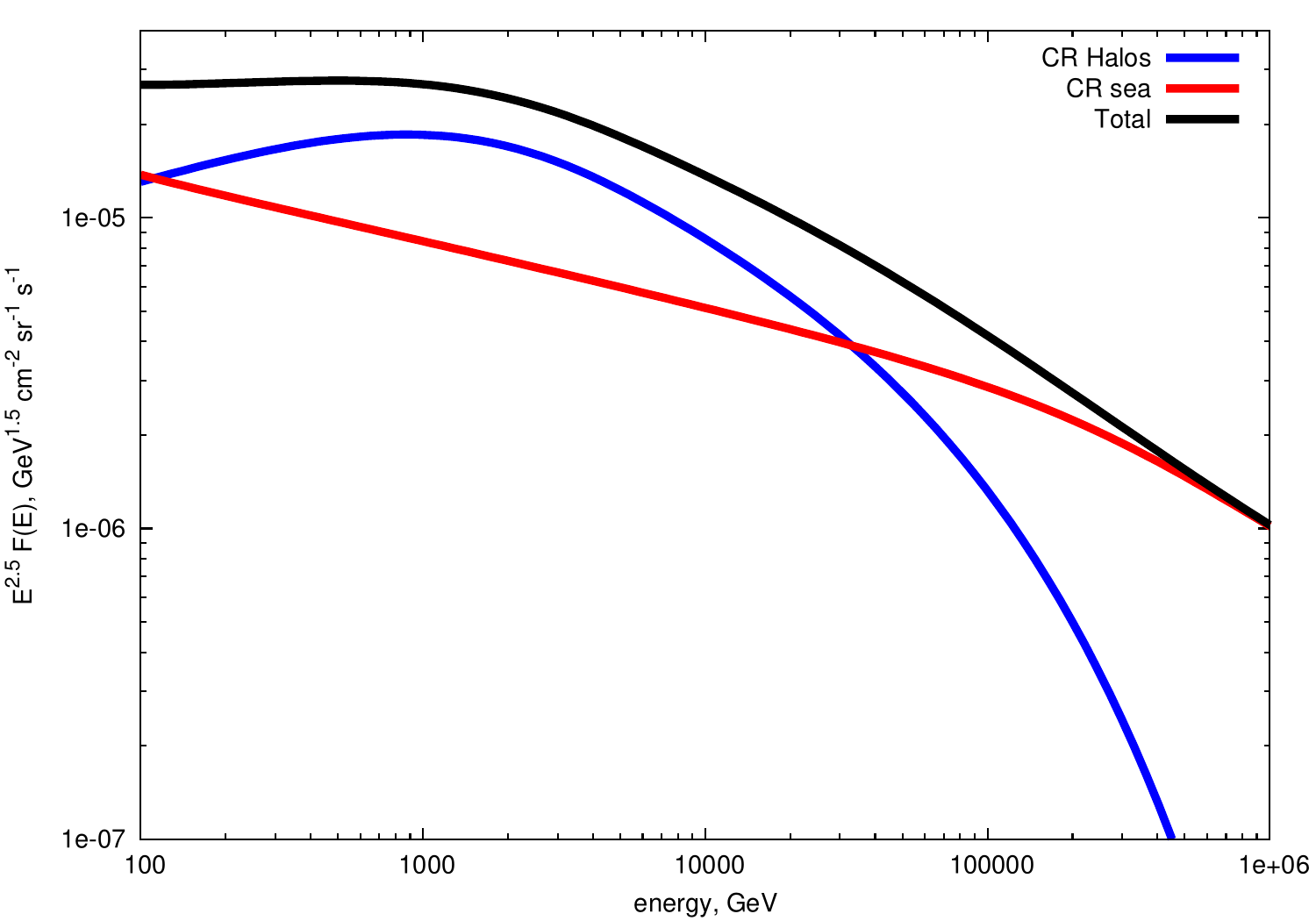}
\caption {
The diffuse all flavor neutrino fluxes in the same region and with the same CR spectrum as used in Fig.\ref{fig:dif}.  
 }
\label{fig:neu}
\end{figure}

Finally, our interpretation of the excess multi-TeV diffuse gamma-ray emission predicts enhanced neutrino fluxes from the slow diffusion regions surrounding the CR accelerators. The expected neutrino fluxes from the inner Galaxy using the parametrization of neutrino production cross sections from ref.\cite{kelner06}, are shown
in Fig.\ref{fig:neu}.


\vfill
\clearpage

\noindent {\bf References}
\\
\\
\bibliographystyle{naturemag}

\bibliography{main}
 
 \section*{Author Contributions}
The authors contributed to the paper in equivalent fractions.

\section*{Data Availability}
The data points from AMS-02, DAMPE and LHAASO experiments are from the corresponding publications cited in this work, respectively. 

\section*{Code Availability}
We used the standard data analysis tools in the Python environments, including methods in Astropy, NumPy, Matplotlib. All these packages are publicly available through the Python Package Index (https://pypi.org).

\section*{Acknowledgments} 
Rui-zhi Yang is supported by the NSFC under grant 12393854 and 12041305. Ruizhi Yang gratefully acknowledge the support of Cyrus Chun Ying Tang Foundations.

\begin{methods}

\section*{calculation of B/C ratio}

The  secondary-to-primary ratio can be expressed as  (see e.g. ref.\cite{katz10}), 
\begin{equation}
R(E)=\frac{\frac{X_{\rm ISM}(E)}{m_p}S(E)/N_p(E)}{1+\sigma_{\rm t}\frac{X_{\rm ISM}(E)}{m_p}}, 
\end{equation}
where
\begin{equation}
S(E)=\int_E\frac{d\sigma_{\rm p \to s}(E,E')}{dE'}N_p(E')dE'.
\end{equation}
Here E is the particle energy per nucleon, $N_p(E)$ is the primary CR spectrum  in the ISM, $X_{\rm ISM }$ is the propagation length in $\rm g/cm^2$ ("grammage"), $\sigma_{\rm p \to s}$ is the  differential cross section  of production of the given secondary particle,  $\sigma_{\rm t}$ is the total destruction cross-section of secondaries.  For the spallation reactions,  the energy per nucleon before and after the reaction remains the same (see, e.g., ref.\cite{webber03}). Thus Eq.(1) can be reduced to
\begin{equation}
R(E)=\frac{\frac{X_{\rm ISM}(E)}{m_p}\sigma_{p \to s}}{1+\sigma_{\rm t}\frac{X_{\rm ISM}(E)}{m_p}} . 
\end{equation}
In the calculation, the cross-sections of nuclear reaction are taken from ref.\cite{webber03}. For the production of the 
For the nuclei spallation reactions, we use the parametrizations of cross-sections from ref.\cite{letaw83}. 

The "grammage" accumulated in the ISM is proportional to the confinement time of CRs inside the Galaxy  $\tau(E)$ ,  
\begin{equation}
 X_{\rm ISM}(E) = c~n_{ism}~m_p~\tau_{}(E), 
\end{equation}
where  $c$ is the speed of light,  and $m_p$ is the  proton mass.
We present  $X_{\rm ISM}(E)$ in the form 
\begin{equation}
X_{\rm ISM}(E)=X_0 (10~\rm GeV)^{-\delta} \ .
\end{equation}
Note that $X_{\rm ISM}$ depends on rigidity, $R=\frac{AE}{Ze}$, where $A$ and $Z$ are  the atomic and mass number, respectively. For  
protons and antiprotons,  R=E, while for nuclei it differs by a factor of $\approx 2$.  

For the grammage accumulated  in the slow diffusion `bubble' near the source,  an  additional component of secondary antiparticles should be invoked: 
\begin{equation}
R(E)=R1(E)+R2(E) .
\end{equation}
The ratio R1(E) is for secondaries produced in the ISM ($X_{\rm ISM}$); see  Eq.(7). The ratio  R2(E) corresponds  to the second component represented by secondary particles produced inside  the slow diffusion `bubbles' near the source ($X_{\rm source}$)  
\begin{equation}
R2(E)=\frac{\frac{X_{\rm s}(E)}{m_p}S'(E)/N_p(E)}{(1+\sigma_{s}\frac{X_{\rm ISM}(E)}{m_p})(1+\sigma_{s}\frac{X_{\rm s}(E)}{m_p})},
\label{eq:r2}
\end{equation}
where (for spallation)
\begin{equation}
S'(E)=d\sigma_{p-s}(E) Q_{s}(E) \tau(E).
\end{equation}
Here $Q_{s}(E)$ is the primary injection spectrum from the source with a power law index $\gamma_{\rm s}$,  $X_{\rm source}$ is the ??grammage?? accumulated inside the slow diffusion `bubble' near the source. We represent $X_{\rm s}$ in the form of broken power law as described in the text.

\section*{Gamma-ray emission from Cygnus region}
Cygnus-X, an intensive nearby star-forming region, is one of the brightest regions in the gamma-ray band. At GeV energies, Fermi-LAT has revealed an extended gamma-ray structure with an extension of more than 2 degrees (in radius) and a hard spectrum (photon index of 2.2) \cite{fermi_cygnus}, dubbed as Cygnus cocoon. Later, HAWC collaboration reported the detection of the Cygnus Cocoon in the TeV band.   Recently, LHAASO reported a much more extended gamma-ray  bubble beyond the cocoon \cite{lhaaso_cygnus}. The LHAASO analysis revealed that the gamma-ray spectrum in the extended 'bubble` is identical to the inner `cocoon'.
In contrast, the non-detection of 'bubble` in other energy bands is presumably caused by the lower sensitivities of the detectors in comparison with LHAASO. Indeed, the recent Fermi LAT results on Cygnus region reveal a more extended gamma-ray  emitting region \cite{astiasarain23}. Since there is no detection of Cygnus Bubble at low energies, we use the Fermi-LAT and  HAWC gamma-ray spectra in the cocoon and the LHAASO-KM2A gamma-ray spectrum in the  `inner bubble' as a representative of gamma-ray  spectrum in the vicinity of CR accelerator.  These components are consistent in morphology, i.e., they can all be described as a Gaussian with $\sigma\sim 2^{\circ}$.  


\clearpage

%
%

\renewcommand{\refname}{References}

\bibliographystyle{naturemag}

\end{methods}
\end{document}